\documentclass[useAMS,usenatbib]{mn2e} \input{psfig}

\title[Radio and optical orientations of galaxies]
  {Radio and optical orientations of galaxies}
\author[Battye \& Browne]
  {R.A.~Battye$^1$, I.W.A.~Browne$^1$\\
    $1$Jodrell Bank Centre for Astrophysics, School of Physics and Astronomy, The University of Manchester, Oxford Road, Manchester, M13 9PL}
\date{Released 2009 September 1}

\pagerange{\pageref{firstpage}--\pageref{lastpage}} \pubyear{2009}

\def\LaTeX{L\kern-.36em\raise.3ex\hbox{a}\kern-.15em
    T\kern-.1667em\lower.7ex\hbox{E}\kern-.125emX}

\begin{document}

\label{firstpage}

\maketitle

\begin{abstract}
We investigate the correlations between optical and radio isophotal
position angles for 14302 SDSS galaxies with $r$ magnitudes brighter
than 18 and which have been associated with extended FIRST radio
sources. We identify two separate populations of galaxies using the
colour, concentration and their principal components.  Surprisingly
strong statistical alignments are found: late-type galaxies are
overwhelmingly biased towards a position angle differences of
$0^{\circ}$ and early-type galaxies to $90^{\circ}$.  The late-type
alignment can be easily understood in terms of the standard picture in
which the radio emission is intimately related to areas of recent
star-formation.  In early-type galaxies the radio emission is expected
to be driven by accretion on to a nuclear black hole. We argue that
the observed correlation of the radio axis with the minor axis of the
large-scale stellar distribution gives a fundamental insight into the
structure of elliptical galaxies, for example, whether or not the
nuclear kinematics are decoupled form the rest of the galaxy. Our
results imply that the galaxies are oblate spheroids with their radio
emission aligned with the minor axis.  Remarkably the strength of the
correlation of the radio major axis with the optical minor axis
depends on radio loudness. Those objects with a low ratio of FIRST
radio flux density to total stellar light show a strong minor axis
correlation while the stronger radio sources do not. This may reflect
different formation histories for the different objects and we suggest
we may be seeing the different behaviour of slow rotating and
fast rotating ellipticals.  A simple analysis to estimate
the effects of measurement errors indicates that the intrinsic degree
of anti-alignment in the roundest early-type galaxies may be as small
as $\pm 15^{\circ}$ and a similar value is obtained for the degree of
alignment in the late-type population.


\end{abstract}

\begin{keywords}
 Galaxies: -- radio galaxies -- star-forming galaxies.
\end{keywords}

\section{Introduction}
There has been a long history of searching for alignments between the
structures of radio sources and those of their host galaxies and
generally the results have been inconclusive (for example,
\cite{Mackay1971}; \cite{Palimaka1979}; \cite{Valtonen1983};
\cite{Birkinshaw1985}; \cite{Sansom1987}), though in most
investigations there appear to be a preponderance of objects where the
radio elongation is more aligned with the optical minor axis than the
major axis. The clearest result was obtained by \cite{Condon1991}
who found that extended radio jets in 125 UGC galaxies were
preferentially aligned with the optical minor axes of their hosts,
with the effect being strongest for elliptical galaxies. Other,
related investigations, have focused on the relative orientation of
radio axes and dust disks (for example, \cite{Schmitt2002}) and
these again show a broad distribution with no obvious preferred
alignment. Such studies were motivated by the desire to find a
connection between radio emission mechanism and the geometry of the
host galaxy. The historic results have been based on relatively small
numbers of objects ($\sim$ 100) but with the advent of deep radio and
optical surveys like Faint Images of the Radio Sky at Twenty
centimetres (FIRST; \cite{Becker1995}) and the Sloan Digital Sky Survey (SDSS; \cite{York2000}),
respectively, it is possible to construct samples with orders of
magnitude larger numbers. This means going deeper in radio flux
density to get better statistics and an unavoidable consequence is
that the galaxies identified with the radio sources are no longer
dominated by the ellipticals, which were the targets of early studies,
but are now a mixture of ellipticals and disk-dominated star-forming
galaxies. Since the source of the emission is fundamentally different
in the two types of galaxy one might well expect their alignment
properties also to be different. For example, it is known from studies
of nearby star-forming galaxies that the radio emission traces the
distribution of recent star-formation (\cite{Condon1992} and references
therein; \cite{Murphy2008}) and therefore, assuming one has
observations with the appropriate resolution, one might expect to see
a statistical alignment between the radio emission and the
starlight. On the other hand radio emission in the elliptical
population is likely to be powered by the nuclear black hole and the
alignment of the resulting jets/lobes will contain information about
the nucleus of the galaxy. In the light of the existence of kinematic
sub-structures, including decoupled cores, in a significant fraction
of early-type galaxies (for example, \cite{Halliday2001}; \cite{Loubser2008}; \cite{Krajnovic2008}), and the separation of such
galaxies into two distinct classes the fast and slow rotators
(\cite{Emsellem2007}), studying the alignment between the stellar
population and the radio emission assumes added significance.

In this paper we examine the statistics of the alignments of radio and
optical structures using radio data from the FIRST survey (\cite{Becker1995}) which has a resolution $5^{\prime\prime}$ and optical data
for the identifications from SDSS (\cite{York2000}).  Star
formation, and its associated radio emission, tends to be concentrated
in the inner few kpc of galaxies (for example, \cite{Condon1992}; \cite{Muxlow2005}; \cite{Wilman2008}). Thus one might expect such emission to
be just resolved in most FIRST radio maps of star-forming galaxies
with low redshift ($\leq$ 0.2). In such galaxies optical elongations
should be available in SDSS and thus the FIRST/SDSS combination is
well-suited to the task. For the other major component of the mJy
radio population, the jet-powered radio galaxies, FIRST data are also
useful but for a more restricted set of objects because of the wide
dispersion in their linear sizes. The great advantage of using SDSS is
the wealth of photometric and spectroscopic data that can be used to
separate intrinsically different populations of galaxies and we
exploit this extensively in what follows.

Our motivations for this work are to learn more about the mechanisms
behind the alignments and to elucidate the statistical properties of
the low luminosity radio emitters which will dominate the populations
of radio sources discovered in new deeper radio surveys like LOw
 Frequency ARray (LOFAR),
Square Kilometre Array (SKA) pathfinders 
and the SKA itself. There is the possibility that the
extended radio emission from star-forming galaxies might be used for
weak lensing studies (\cite{Blake2004}) and it is interesting to
investigate properties of the observed galaxy alignments when compared
to the optical.

\section{Sample selection}

We use optical identifications of FIRST radio sources listed in the
SDSS DR6 database (\cite{Adelman2008}) which were selected
using the procedure defined in \cite{Ivezic2002}.  There are
239993 such identifications which were extracted from the SDSS
database using SQL. We have also extracted the photometric model
magnitudes from SDSS ($ugriz$), the integrated flux density measured
by FIRST ($S_{\rm int}$), the position angle (PA, $\alpha$ defined
between $-90^{\circ}$ and $90^{\circ}$, and measured from North
through East) computed from the isophotal distribution in the $r$-band
by SDSS along with the equivalent from FIRST, the major ($a$) and
minor ($b$) axes measured by SDSS and FIRST, and $R_{50}$ and $R_{90}$
measured from the Petrosian intensity profile model fitted to the SDSS
images (\cite{Petrosian1976}). The quantity $c=R_{90}/R_{50}$ is known as
the concentration; for $c$ small the light distribution has an
exponential profile and fits the profile of a disk galaxy, whereas for
large values of $c$ the profile is approximated by a de Vaucouleurs
profile which is often used to model the light distribution of an
elliptical galaxy.

For our investigation we require galaxies and radio sources for which
there are reliable measures of the PAs of extended emission. In order
to produce a statistically robust sample we have imposed additional
selection criteria. First, we have excluded all galaxies which are
classified as ``stellar'' by the SDSS pipeline. Such galaxies are likely
to be at high redshift and will be almost circular from the point
of view of the SDSS beam, making computation of a meaningful PA
impossible. It will also be difficult to compute the PA accurately for
very faint galaxies, those with large axial ratios $(b/a)$ and those
where the major axes are significantly smaller than the beam. We have
excluded all galaxies with $r>18$, $b/a>0.8$ in either SDSS or FIRST,
and those with $a<2^{\prime\prime}$ in FIRST\footnote{The FIRST
beamwidth is $5^{\prime\prime}$ but deconvolved sizes with
$a<1^{\prime\prime}$ are listed.}. This leaves a total of 14302
galaxies on which we have performed our analysis. The size of this
sample is around 2 orders of magnitude larger than previous studies
and it contains a range of galaxy types.

\begin{figure}
\centerline{\psfig{figure=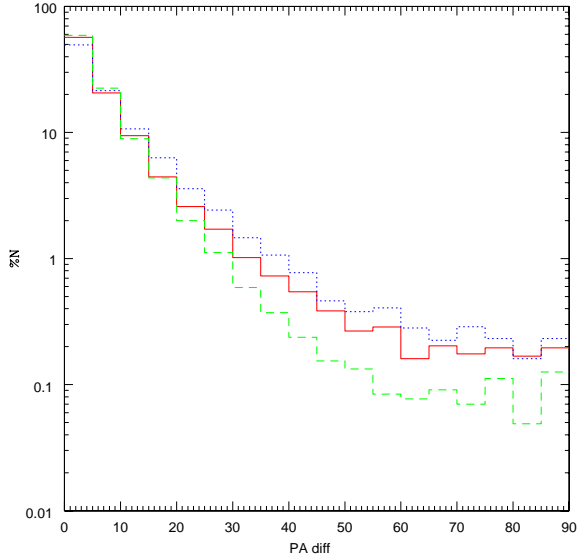,width=8cm,height=8cm}}
\caption{Comparison of the PA measured in different bands. The
histograms are of the percentage of galaxies in each of the bins. The
solid line represents the comparison between $r$ and $g$, the dotted
line that between $g$ and $i$, and the dashed between $i$ and $r$. In
all three cases the histograms are peaked at zero and the differences
in PA are less than $20^{\circ}$ for $\geq$96\% of the galaxies.}
\label{fig:angcomp}
\end{figure}

\begin{figure}
\centerline{\psfig{figure=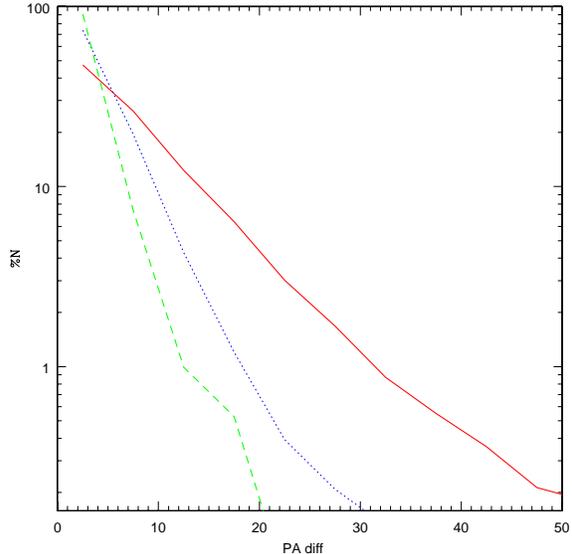,width=8cm,height=8cm}}
\caption{The percentage of galaxies against the PA
difference measured in the $r$ and $u$ bands as a function of axial
ratio $b/a$. The solid line has $0.6<b/a<0.8$, dotted line
$0.4<b/a<0.6$ and dashed line $b/a<0.4$. The equivalent histogram for
all values of the axial ratio is the dashed line in
Fig.~\ref{fig:angcomp}.}
\label{fig:angcomp_ratio}
\end{figure}

We have decided to use the $r$-band PAs since they are computed from
the images which typically have the highest signal-to-noise. In
Fig.~\ref{fig:angcomp} we present histograms of the acute angle
between the PAs measured in the next two most sensitive bands, $i$ and
$g$, for the sample with $r<18$. It can be seen that there is very good
agreement between them. Relaxation of the limit on $r$ degrades this
agreement, whereas imposing a more severe cut-off would produce an
even tighter correlation between the PAs measured in different
bands. We have also plotted the difference between the $r$ and $i$
band PAs as a function of $b/a$ in Fig.~\ref{fig:angcomp_ratio}. It
shows that, as one would expect, the noise on measurement of the PAs
is lower for objects with much better defined PAs, that is, those
which have small axial ratios.

\section{Results}

\subsection{Whole sample}

\begin{figure}
\centerline{\psfig{figure=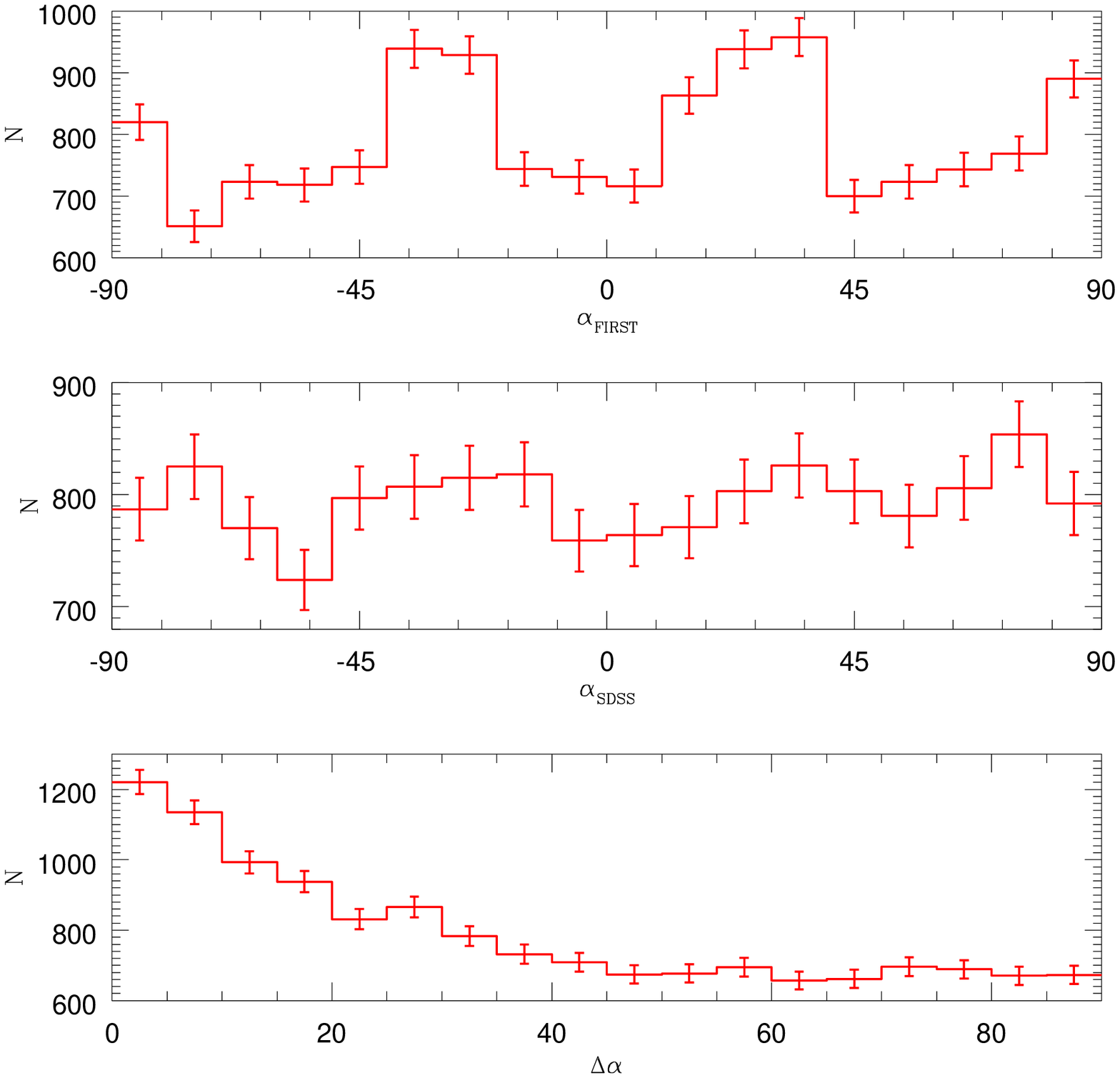,width=8cm,height=12cm}}
\caption{Histograms of $\Delta\alpha$ (bottom), $\alpha_{\rm SDSS}$
(middle) and $\alpha_{\rm FIRST}$ (top) using 18 bins for all 14302
galaxies. Included also are the Poisson errors for each bin. Note the
peak at $\Delta\alpha=0$ and those at $\alpha_{\rm FIRST}\approx \pm
30^{\circ}$ and $\pm 90^{\circ}$. The histogram for $\alpha_{\rm
SDSS}$ is consistent with a uniform distribution.}
\label{fig:histall}
\end{figure}

\begin{figure}
\centerline{\psfig{figure=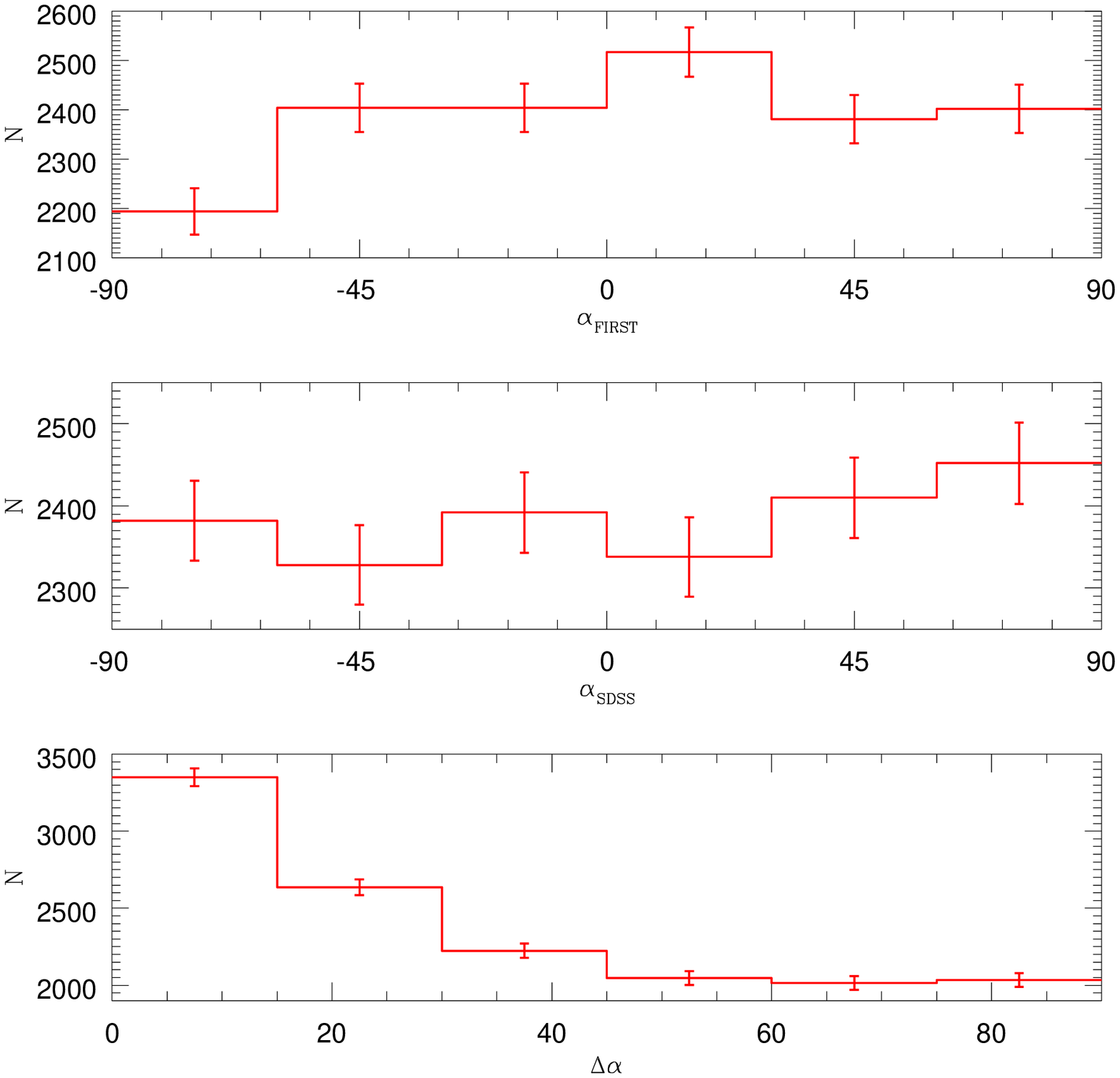,width=8cm,height=12cm}}
\caption{Equivalent to Fig.~\ref{fig:histall} but for 6 bins. The peak at $\Delta\alpha=0$ remains but those in $\alpha_{\rm \small FIRST}$ are smeared out. The distribution of $\alpha_{\rm \small FIRST}$ is still not compatible with a uniform distribution.}
\label{fig:histall6}
\end{figure}

In what follows we will present histograms of the galaxies in the
sample binned according to the quantity $\Delta\alpha$, defined to be
the acute angle between the two position angles $\alpha_{\rm \small
SDSS}$ and $\alpha_{\rm \small FIRST}$. The results for a histogram of
18 bins are presented in Fig.~\ref{fig:histall}, along with the
individual distributions for $\alpha_{\rm \small SDSS}$ and
$\alpha_{\rm \small FIRST}$.  There is a clear excess of galaxies in
which the major axes of the radio and optical emission are aligned,
$\Delta\alpha=0^{\circ}$. Comparing the observed distribution with a
uniform one in which bin contains the average number per bin of the
observed distribution, we obtain $\hat\chi^2(=\chi^{2}/\hbox{degree of
freedom})\approx 37$. The number of degrees of freedom is the number of bins
minus one since we leave the mean free.
This can be increased to $\approx 117$ using 6
bins as illustrated in Fig.~\ref{fig:histall6}. We will use 6 bins as
the standard in what follows.

The histogram of $\alpha_{\rm \small SDSS}$ is compatible with a
uniform distribution $\hat\chi^2\approx 1.1$. By contrast that for
$\alpha_{\rm \small FIRST}$ appears to have statistically significant
peaks at $\pm 30^{\circ}$ and to a lesser extent at $\pm
90^{\circ}$. The peaks are clearly systematics associated with the extraction
of PAs from the images produced the FIRST survey. 
\cite{Chang2004} have searched for cosmic shear signal due to weak
gravitational lensing using the data from the FIRST survey. They
pointed out a number of subtle systematic effects which had to be
removed from the images in order to make an accurate measurement of
cosmic shear. These effects are small at the level of an individual
source, but since they are systematic they can be seen in global
statistics such as those used to study cosmic shear.  They also show
up in the histogram presented in Fig.~\ref{fig:histall}. However, our
interest is in position angle differences, $\Delta\alpha$, and for
systematics in one set of angles to have a significant effect they
would have to be correlated with systematics in the other set of
angles. Since the SDSS angle distribution is nearly uniform it is
unlikely that such systematics will have a significant effect on our
results. Moreover, as we will see later, some of our most interesting
results are that the observed position angle distributions depend
strongly on the physical properties of the objects, and this cannot
arise from measurement systematics. Nevertheless, to reassure
ourselves of the robustness of our results we have performed some
tests. We have selected a subset of FIRST sources to have a completely
flat distribution of position angles and we have also divided the original
sample into different areas and looked for different systematics in
different areas of sky. We still see similar biases to those in
Fig.~\ref{fig:histall} in the position angle differences in the sample
in which the FIRST position angle distribution is flat. In the second
test we find that the FIRST position angle distribution is roughly the same
independent of where we look in the sky.

\subsection{Split on colour and concentration}

\begin{figure}
\centerline{\psfig{figure=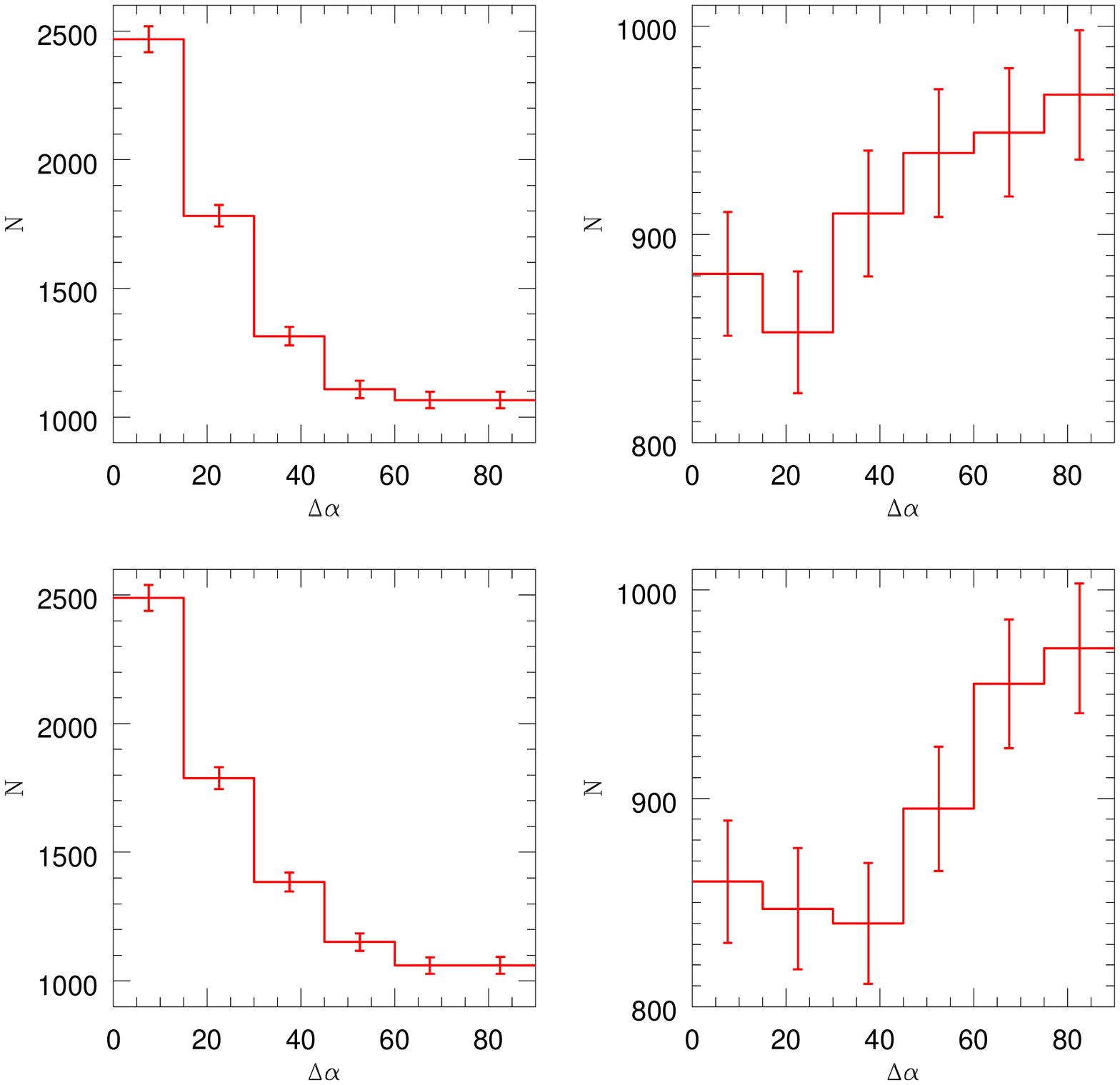,width=8cm,height=8cm}}
\caption{Split of the sample by colour (bottom) and concentration
(top). For colour we split $u-r<3$ (left) and $u-r>3$ (right), and for
concentration we use $c<3$ (left) and $c>3$ (right). The two right-hand
plots are peaked at $\Delta\alpha=0^{\circ}$, whereas the two left
hand plots are biased toward $\Delta\alpha=90^{\circ}$.}
\label{fig:histsplit1}
\end{figure}

It is natural to split the sample based on photometric determined
parameters such as colour and concentration. It has been shown by
\cite{Strateva2001} that the colour defined by $u-r$ can
discriminate between different distributions in the bimodal $g-r$
vs. $u-g$ colour-colour diagram for an earlier SDSS data
release. Based on morphological identifications of nearby galaxies
they suggest that $u-r=2.22$ is the optimal colour separator and then
go on to show this split can be linked directly to two populations,
early-type galaxies (E, S0, Sa) which have $u-r>2.22$ and late-type
galaxies (Sb, Sc, Irr), with $u-r<2.22$. We have examined the
properties of the correlation between $\alpha_{\rm \small SDSS}$ and
$\alpha_{\rm \small FIRST}$ as a function of $u-r$. Specifically we
constructed histograms as a function of $u-r$ with bin width
$\Delta(u-r)=0.5$ and searched for obvious trends. We find that the
peak in the histogram at $\Delta\alpha=0^{\circ}$ is apparent for low
values of $u-r$ (blue objects), whereas the histograms for higher
values of $u-r$ (red objects) are peaked at $\Delta\alpha=90^{\circ}$
(correlation between the SDSS major axis and the FIRST minor axis, or
vice-versa). In Fig.~\ref{fig:histsplit1} we have split the sample
into sub-samples with $u-r<3$, 8933 galaxies, and $u-r>3$, 5369
galaxies. This appears to be close to optimal in terms of separating the
two effects. For $u-r<3$ we have that $\hat\chi^2\approx 212$, whereas
$\hat\chi^2\approx 3.6$ for $u-r>3$. Clearly the observed correlation
for blue half of the galaxy sample is extremely strong, but that for
red half is also statistically significant. The precise cut which we
have used is interesting since \cite{Strateva2001} showed that for
$u-r>3$ the sample would be dominated by elliptical morphologies
(including S0s) with little contamination from spirals.  We have also
performed a split on concentration which as discussed in \cite{Strateva2001} can also crudely distinguish between different galaxy
types. Using a similar approach to the case of colour, we have
identified $c=3$ as being close to the optimal in separating the two
effects already noted above. Fig.~\ref{fig:histsplit1} shows
histograms for $c<3$, 8803 galaxies, and $c>3$, 5499 galaxies. The two
samples have $\hat\chi^2\approx 215$ and 2 respectively. The peak at
$\Delta\alpha=90^{\circ}$ for $c>3$ is less significant than the
corresponding one when colour selection is used, suggesting that there
is a little more cross-contamination of the sample when using
concentration as a discriminator.

On the basis of these two simple splits we are already led to the
basic conclusions of this work: blue, less concentrated galaxies,
which broadly represent the disk-dominated, spiral population, have
the optical and radio major axes correlated, whereas at least some
fraction of red, highly concentrated galaxies, which are part of the
elliptical population, have a correlation between the optical major
and radio minor axes. In the subsequent sections we will attempt to
refine the selection process using other measured properties of these
galaxies. The aim is to strengthen the correlations as quantified by
the value of $\hat\chi^2$ and to obtain a physical insight as to their
origin.

\subsection{Principal component analysis}

\begin{figure}
\centerline{\psfig{figure=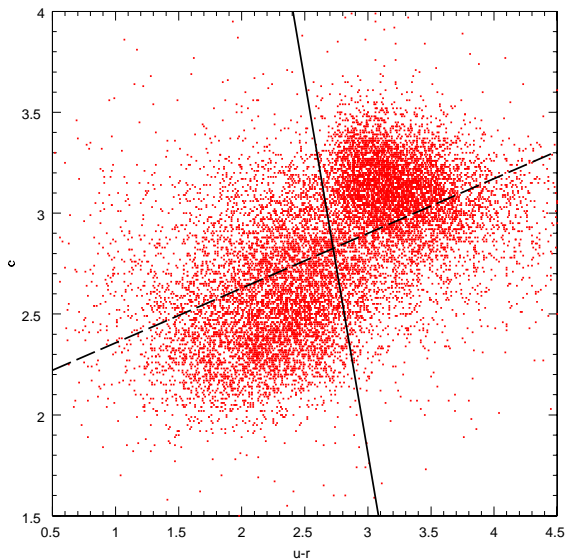,width=8cm,height=8cm}}
\caption{Scatter plot of the colour $u-r$ vs. $c$ for
the whole sample of 14302 galaxies. Included also are the lines
$C_1=2.0$ (dashed) and $C_2=3.5$ (solid).}
\label{fig:colvcon}
\end{figure}

The colour and concentration are known to be correlated (see for
example, \cite{Strateva2001}). In Fig.~\ref{fig:colvcon} we
illustrate this for our sample by presenting a scatter plot of $u-r$ and
$c$ which clearly shows the degeneracy between these two
properties. There is also a bimodality in the density of points in
this plot, suggesting that there are indeed two populations within the
sample.

In order to investigate this quantitatively we have performed a principal component analysis (PCA) on the sample, initially with two variables $u-r$ and $c$. The two components which are generated by this procedure are 
\begin{eqnarray}
C_1&=&0.965c-0.262(u-r)\,,\cr
C_2&=&0.262c+0.965(u-r)\,.
\end{eqnarray}
$C_1$ has the lowest standard deviation, 0.33, and that for $C_2$ is
0.81. A line corresponding to $C_1=2.0$, which is its mean, is
presented in Fig.~\ref{fig:colvcon}. If one considers the rotated
coordinate system $(C_1,C_2)$, the degeneracy obvious in
Fig.~\ref{fig:colvcon} is along the $C_2$ axis and therefore the value
of $C_2$ can be used to quantify the position along the
degeneracy. Given that the observed bimodality appears to be
compatible with the split along the degenerate line, we have created a
set of histograms which split the sample in terms of the value of
$C_2$ and we find that $C_2=3.5$ is close to optimal in terms of
separating the two populations.

\begin{figure}
\centerline{\psfig{figure=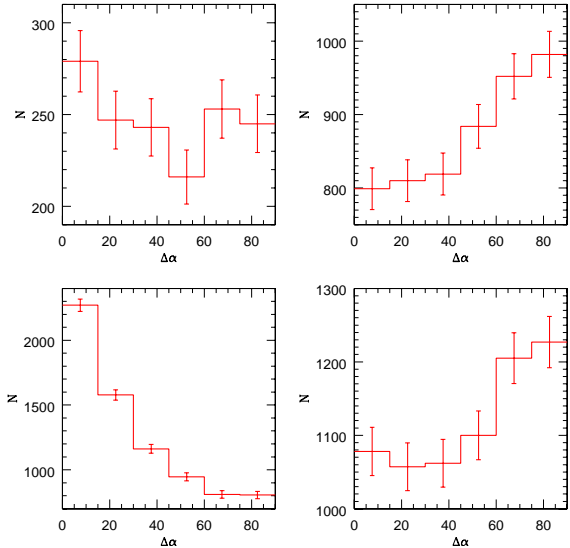,width=8cm,height=8cm}}
\caption{Various splits of the data determined on the basis of the PCA analysis and the axial ratio: (bottom left) $C_2<3.5$; (bottom right) $C_2>3.5$; (top left) $C_2>3.5$ and $b/a<0.6$; (top right) $C_2>3.5$ and $b/a>0.6$.}
\label{fig:histcorr}
\end{figure}

In the bottom two panels of  Fig.~\ref{fig:histcorr} we present results of splitting the sample using $C_2$. There are
7573 galaxies with $C_2<3.5$, the histogram is biased toward
$\Delta\alpha=0^{\circ}$ and $\hat\chi^2\approx 260$ improving the
correlation from the simpler splits $u-r<3$ and $c<3$. When one
creates a sub-sample with $C_2>3.5$ there are 6729 galaxies, the bias
is now toward $\Delta\alpha=90^{\circ}$, and $\hat\chi^2\approx 5$
again appearing to improve the previously identified correlation. It seems
therefore that $C_2$ is a better discriminator of the populations
responsible for the two observed correlations than either colour or
concentration by themselves.

We have also considered further splits on the basis of the optical
axial ratios, $b/a$, of the galaxies as measured by SDSS, in addition
to $C_2$. One reason for doing this is that for high axial ratios the
position angle of elongation becomes increasing difficult to measure
reliably. Also \cite{Condon1991} found that minor axis alignments
were strong only for the rounder galaxies. Splitting by axial ratio
appears to have little effect when $C_2<3.5$ (not shown), but for
$C_2>3.5$ the situation is very different as shown in the top two
panels of Fig.~\ref{fig:histcorr} which shows only objects with
$C_2>3.5$; that is the red highly concentrated objects which are
predominantly early-type galaxies. For $b/a>0.6$ there are 5246
galaxies and there appears to be a bias toward
$\Delta\alpha=90^{\circ}$ with $\hat\chi^2\approx 7$, whereas for
$b/a<0.6$ there are 1483 galaxies with a (less significant) bias
toward $\Delta\alpha=0^{\circ}$ with $\hat\chi^2\approx 1.7$. This
suggests that a split based on axial ratio produces cleaner
subsamples. It also shows that measurement errors on optical position
angles are not strongly biasing our results since the stronger
correlation is seen for the more circular galaxies. It is interesting
that splitting by axial ratio seems to separate two populations
amongst these predominantly elliptical galaxies, one with
$\Delta\alpha=90^{\circ}$ and the other with
$\Delta\alpha=0^{\circ}$. We explore this further below when we also
split the population in terms of their radio loudness.

\begin{figure}
\centerline{\psfig{figure=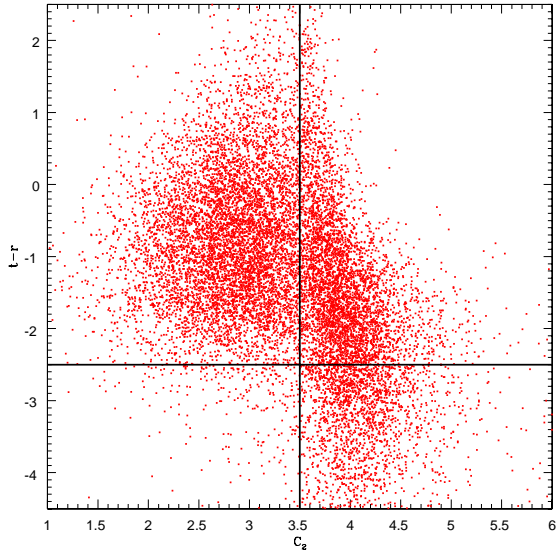,width=8cm,height=8cm}}
\caption{Scatter plot of $t-r$ vs. $C_2$. Included also are lines corresponding to $C_2=3.5$ and $t-r=-2.5$.}
\label{fig:corr1vloud}
\end{figure}

Following \cite{Ivezic2002} it is convenient to define an AB radio
magnitude in terms of the integrated flux density
\begin{equation}
t=-2.5\log_{10}\left(S_{\rm int}\over 3631{\rm Jy}\right)\,,
\end{equation}
and the parameter $t-r$ then quantifies whether the galaxy is
relatively radio-loud or radio-quiet. To put the degree of radio loudness
in context we calculate that $t-r\sim-7$ corresponds to an $L^{*}$
galaxy having a radio luminosity which would put it roughly at the
dividing line between radio sources having FR1 and FR2 radio
structures (\cite{Ledlow1996}). Though the galaxies in our sample
have a range of optical luminosities, the vast majority will have
radio luminosities below the FR1/FR2 division. We note that since
the beamsize of FIRST is only 5 arcseconds, the measured radio flux
density will be that associated with the nuclear region and will
sometimes be an under estimate of the total flux density. This fact
should be borne in mind when interpreting the results presented
here. It also should be made clear that our radio-quiet/loud division
used later in this paper at $t-r=-2.5$ is different from the standard
definition in two ways; we do not use the total radio flux density
and, more importantly, we take the ratio of the radio emission to the
total optical starlight and not the AGN light\footnote{Coincidently
the numerical value for the separation between our radio-loud and
radio-quiet populations is the same as that usually adopted to
separate radio-loud and radio-quiet quasars (\cite{Kellermann1994}}.  In Fig.~\ref{fig:corr1vloud} we present a scatter plot of
the $t-r$ versus $C_2$ which appears to have an interesting
morphology. One can clearly see that there are two populations
separated by $C_2=3.5$. The population with $C_2<3.5$ has
$-2.5<t-r<1.0$, whereas that with $C_2>3.5$ extends over a much wider
range $-4.5<t-r<1.5$.

\begin{figure}
\centerline{\psfig{figure=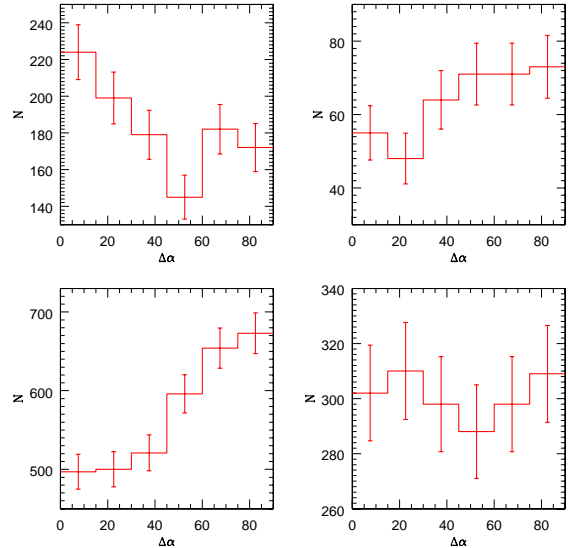,width=8cm,height=8cm}}
\caption{Histograms for  the case $C_2>3.5$ with  various splits based
on $b/a$ and $t-r$. The two plots at the bottom have $b/a>0.6$ and the
two the  top have $b/a<0.6$. The two  on the left are  radio-quiet with
$t-r>-2.5$ and the two on the right are radio-loud with $t-r<-2.5$.}
\label{fig:histsplit2}
\end{figure}

In Fig.~\ref{fig:histsplit2} we have split the subsample with
$C_2>3.5$ based on both $b/a$ and $t-r$. For $b/a>0.6$ and $t-r>-2.5$
(radio-quiet; bottom left panel) there are 3441 galaxies. Their
histogram is biased toward $\Delta\alpha=90^{\circ}$ with
$\hat\chi^2\approx 11$ which is the statistically strongest result we
have been able to find for the correlation between major axis of the
optical image and the minor axis in the radio using just photometric
properties. Remarkably, for the radio louder population  with $b/a>0.6$ the
distribution of the 1805 objects is statistically indistinguishable
from uniform. Thus, amongst these galaxies with $b/a>0.6$ we appear to have
isolated two populations exhibiting different behaviour depending on
their degree of radio-loudness.  We think that the dichotomy between
radio-quiet and radio-loud galaxies for $b/a>0.6$ is an intriguing
result worthy of follow-up. We will discuss this further in Section 4.

As we shall explain in section 3.4, we believe that the apparent trend
towards $\Delta\alpha=0^{\circ}$ in the radio quieter sub-sample with
$b/a<0.6$ (top left of Fig.~\ref{fig:histcorr}) is due to
contamination of the elliptical population by red star-forming
galaxies. The $b/a<0.6$ histogram for the radio louder objects (top
right of Fig.~\ref{fig:histcorr}), which contains 382 galaxies, is
biased toward $\Delta\alpha=90^{\circ}$, albeit with a low
$\hat\chi^2\approx 1.6$. We deem this to be compatible with
uniformity.

\subsection{Spectroscopic identifications}

\begin{figure}
\centerline{\psfig{figure=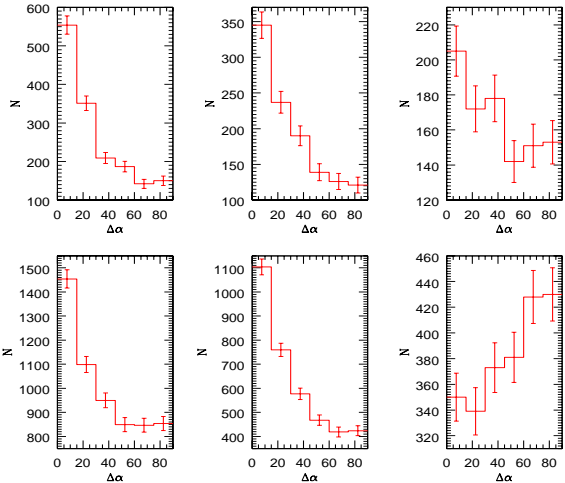,width=8cm,height=7cm}}
\caption{Histograms of the sample split on the basis of spectroscopic
data: (bottom left) all galaxies with spectroscopic data; (bottom
middle) those classified on the basis of spectral lines
from Brinchmann et al (2004); (bottom right) those with spectroscopic
data, but were not classified on the basis of spectral lines; (top
left) those classified as star-forming galaxies; (top middle) those
classified as composite; (top right) those identified as AGN.}
\label{fig:histspec}
\end{figure}

\begin{figure}
\centerline{\psfig{figure=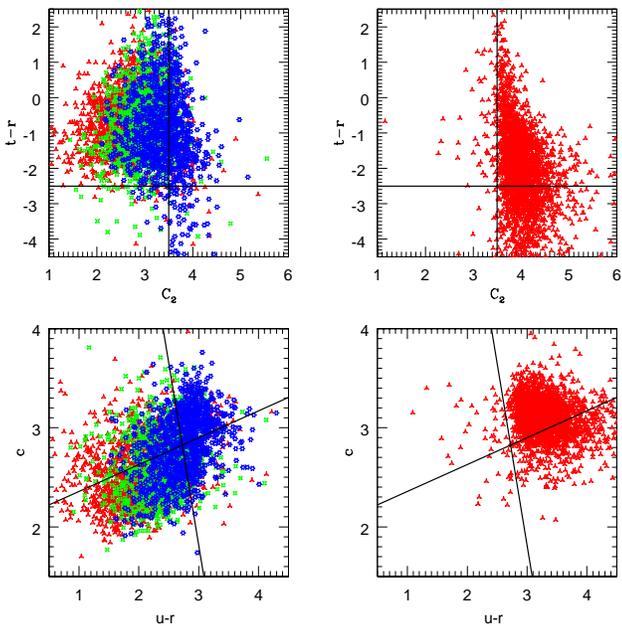,width=9cm,height=9cm}}
\caption{Scatter plot of $c$ vs. $u-r$ (bottom) and $t-r$ vs. $C_2$
(top). Only objects whose spectra classify them as star-forming
(triangles), composite (squares) or AGN (hexagons) are in the left
hand plots, whereas those which cannot be classified on the basis of
their spectra are on the right. Included also are lines presented in
Figs.~\ref{fig:colvcon} and \ref{fig:corr1vloud}. Note that the splits
$C_2<3.5$ and $C_2>3.5$, appears to be virtually the same as the split into
classified and unclassified, respectively.}
\label{fig:split}
\end{figure}

In the discussion so far we have relied on galaxy properties derived
from photometric data. In addition to photometry many of the galaxies
have SDSS spectra. These data have been analysed to compute the
redshift of the object, but they have also been used by a number of
authors to derive a range of galaxy properties such as the stellar
mass, age and star-formation rate. In particular, \cite{Kauffmann2003} used SDSS spectroscopy to compute the
stellar mass, $m_{\rm stellar}$, and age, $t_{\rm stellar}$, from two
stellar absorption line ratios coupled to broadband photometry, while
\cite{Brinchmann2004} list star-formation rates
(SFRs) and other parameters for SDSS galaxies using similar
techniques\footnote{see
http://www.mpa-garching.mpg.de/SDSS/DR4/Data/sfr catalogue.html for
access to these data.}. We note that these calculations use the DR4
catalogue whereas we have used DR6. We have used the SDSS identifiers
to match up the two catalogues. For the objects we have selected, all
positions in DR4 are consistent with those in DR6.

\cite{Brinchmann2004} also attempt a classification of the galaxies
on the basis of their spectral lines, particularly on the location of
objects in the plane defined by ratios [NII]/H$\alpha$ and
[OIII]/H$\beta$; that is where objects lie in the
Baldwin-Phillips-Terlevich (BPT) diagram, following \cite{Baldwin1981}. The galaxies are divided into four groups: star-forming,
active galactic nuclei (AGN), composite (that is, have both
star-forming and AGN properties) and unclassified. Before we discuss
the results of our analysis using the SDSS spectroscopic data a word
of caution is appropriate; the spectra were obtained with a 3~arcsec
fibre bundle and, particularly for low redshift, face-on, galaxies the
spectra may be missing some star-formation in disks. Some of these
relatively rare objects could be labelled as unclassified (see below)
because the spectroscopy only refers to a bulge of a disk galaxy.

The sample of galaxies which we have been studying contains 6053 galaxies which
have entries in the tables of \cite{Kauffmann2003} and \cite{Brinchmann2004}. Of these 3752 were classified using their spectral line
strengths and 2301 were not classified since they have either faint,
or non-existent lines. This latter sample will also be of interest to
us containing, as it does, mostly early-type galaxies, a sub-set of
the red galaxies previously selected by their photometric
properties. Of the spectroscopic sample containing 6053 galaxies, 1593
were classified as star-forming, 1158 as composite and 1001 as dominated
by an AGN. In Fig.~\ref{fig:histspec} we have split the sample
according to this classification scheme. We find that the sample of
star-forming and composite galaxies are significantly biased toward
$\Delta\alpha=0^{\circ}$ with $\hat\chi^2\approx 114$ and 97,
respectively. The AGN are also biased toward $\Delta\alpha=0^{\circ}$,
but with a much lower significance $(\hat\chi^2\approx 3.1$). The
objects which are unclassified spectroscopically appear also to be
biased, but this time towards $\Delta\alpha=90^{\circ}$ with
$\hat\chi^2\approx 3.8$

Although the numbers of objects, and hence $\hat\chi^2$, are lower for
this spectroscopic sample, the percentage of objects in each bin for
star-forming and composite galaxies is similar to that for the
photometric split with $C_2<3.5$. Moreover, percentages for the
unclassified galaxies are similar to those for $C_2>3.5$, indicating a
close link between the spectral classification and this photometric
quantity. This is confirmed in Fig.~\ref{fig:split}, where we have
replotted Figs.~\ref{fig:colvcon} and \ref{fig:corr1vloud} but now
separating spectroscopically classified galaxies and spectroscopically
unclassified galaxies on the left and right, respectively.  There are
virtually no unclassified objects with $C_2<3.5$ ($4\%$ contamination) 
and very few star-forming/composite galaxies with $C_2>3.5$
($7\%$ contamination).  Furthermore, nearly
all the classified galaxies, including the AGN, are radio-quiet (that
is, they have $t-r>-2.5$). This reinforces our view that $C_2>3.5$ is
a good a photometric criterion for selecting passive elliptical
galaxies.

\begin{figure}
\centerline{\psfig{figure=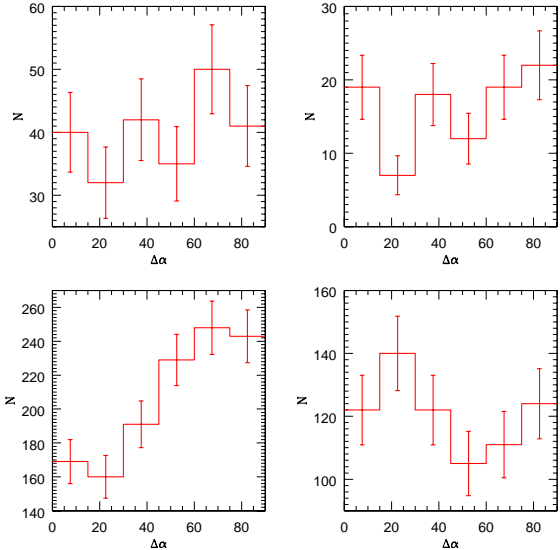,width=8cm,height=8cm}}
\caption{Histograms of the spectroscopically unclassified population (which we
believe are dominated by passive elliptical galaxies). The two plots
at the bottom have $b/a>0.6$ and the two at the top have
$b/a<0.6$. The two on the left are radio-quiet with $t-r>-2.5$ and the
two on the right are radio-loud with $t-r<-2.5$. This plot is
equivalent to Fig.~\ref{fig:histsplit2} except that we have used the
absence of a spectral classification, rather than the condition of
$C_2>3.5$ to pre-select the objects.}
\label{fig:histpass}
\end{figure}

\begin{figure}
\centerline{\psfig{figure=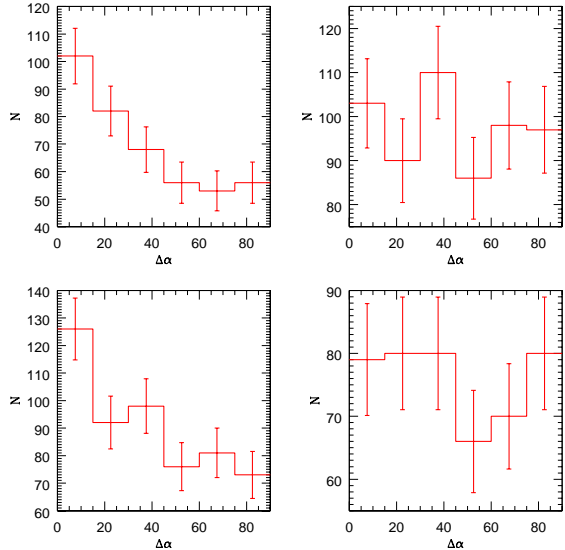,width=8cm,height=8cm}}
\caption{Histograms of splits of the population classified as AGN using their spectra. (bottom left) $C_2<3.5$; (bottom right) $C_2>3.5$; (top left) $b/a<0.6$; (top right) $b/a>0.6$.}
\label{fig:histagn}
\end{figure}

In Fig.~\ref{fig:histpass} we focus on the unclassified population
whose characteristic is that they have no identifiable lines in their
spectra. We believe that this population consists predominantly of
elliptical galaxies in which there is no evidence for current
star-formation nor for the presence of an active nucleus. We will term
these unclassified objects as passive elliptical galaxies. Since they
appear, for the most part, to have $C_2>3.5$ and have a bias toward
$\Delta\alpha=90^{\circ}$, we have attempted to improve the
correlation by making splits into high and low axial ratios, and on
the basis being radio-loud/quiet just as we did in
Fig.~\ref{fig:histsplit2}. For $b/a>0.6$ and $t-r>-2.5$, there are
1240 galaxies with a bias toward $\Delta\alpha=90^{\circ}$ and
$\hat\chi^2\approx 7$. The percentage of galaxies in each bin is
similar to the case of the photometric sample with $C_2>3.5$, although
the value of $\hat\chi^2$ is somewhat lower  due to the reduced
number of galaxies in this sample.  The other three sub-samples
presented in Fig.~\ref{fig:histpass} are $b/a>0.6$ and $t-r<-2.5$,
$b/a<0.6$ and $t-r>-2.5$, $b/a<0.6$ and $t-r<-2.5$ contain 724, 240
and 97 galaxies with $\hat\chi^2\approx 1.20$, 0.96 and 1.91
respectively. All three distributions appear to have no obvious bias
toward a particular angle, and at least in the first two cases are
compatible with uniformity.
 
We briefly return to discuss Fig.~\ref{fig:histsplit2}. We see no
evidence in Fig.~\ref{fig:histpass} for the significant trend seen in
the top left-hand panel of Fig.~\ref{fig:histsplit2}. The availability
of spectroscopic information allows us to understand why there appears
to be a significant non-uniformity for objects in this panel in terms
of a mixture of star-forming galaxies and ellipticals. There are
spectra for 448 galaxies with $C_2>3.5$, $t-r>-2.5$ and $b/a<0.6$. 228
of these are classified as star-forming, composite or AGN and the
histogram of their PA differences peaks at $\Delta\alpha=0^{\circ}$,
while there 240 which are not classified with a peak in their
histogram at $\Delta\alpha=90^{\circ}$. The statistical significance
of the two peaks are much weaker than seen in the rest of the paper,
particularly the one for the unclassified population, due to the
reduced number of galaxies in the sub-sample. However, it seems clear
that objects with $C_2>3.5$ and low values of $b/a$ contain a
significant contamination from star-forming galaxies, composites and
AGN. Visual inspection of the individual images confirms that these
are indeed a population of red ($u-r\sim 3$) spirals. In fact $85\%$
of the galaxies which are spectral unclassified have $b/a>0.6$ (note
that we have excluded galaxies with $b/a<0.8$ from our analysis since
their PA will be affected by noise).  We can therefore refine our
photometric criterion to select a clean sample of passive ellipticals
to include only galaxies with $C_2>3.5$ and $b/a>0.6$.

The AGN population also deserves special mention. Not only is the bias
toward $\Delta\alpha=0^{\circ}$ much weaker than for star-forming or
for composite galaxies, they straddle the line $C_2=3.5$
with almost equal numbers above and below, i.e. equal numbers of star-forming and elliptical galaxies (see \cite{Kauffmann2008a}. In Fig.~\ref{fig:histagn}
we present sub-samples of the AGN population split using $C_2$ and
$b/a$. For $C_2<3.5$ there are 546 AGN which appear to be biased
toward $\Delta\alpha=0^{\circ}$ with $\hat\chi^2\approx 4.2$, whereas
for $C_2>3.5$ there are 455 which appear to be compatible with a
uniform distribution. There is similar dichotomy when one splits
according to $b/a$. There are 417 AGN with $b/a<0.6$ which are biased
toward $\Delta\alpha=0^{\circ}$ with $\hat\chi^2\approx 5.2$ and 584
with $b/a>0.6$ which are a compatible with a uniform distribution. It
seems clear that dividing the sample according to the values of $C_2$
or axial ratio sharpens the correlation toward
$\Delta\alpha=0^{\circ}$ for AGN. Conditions on the axial ratio and
$C_2$ can be coupled together to create sub-samples with less
contamination. There are 279 AGN with $C_2<3.5$ and $b/a<0.6$. Their
histogram has $\hat\chi^2\approx 5.2$ and around 50\% have
$\Delta\alpha<30^{\circ}$. There are 295 AGN with $C_2>3.5$ and
$b/a>0.6$ whose histogram is consistent with a uniform distribution.
It is not clear how to interpret the uniform histograms: it could be
that the samples are genuinely uniform, with no preferred orientation
between the radio and optical axes or, for example, there could be two
populations with roughly equal numbers, biased toward
$\Delta\alpha=0^{\circ}$ and $90^{\circ}$%

\subsection{Physical properties of the aligned galaxy populations}

So far we have used the spectroscopic information to classify (or not)
the galaxies as star-forming, composite, AGN and
unclassified. However, we have already alluded to the fact that
\cite{Kauffmann2003} and \cite{Brinchmann2004}  deduce $m_{\rm
stellar}$, $t_{\rm stellar}$ and the SFR for a significant fraction of
the galaxies we are studying. We therefore have an opportunity to
investigate the connections between these inferred properties of
the galaxies and the correlations we have so far found in this paper.

\begin{figure}
\centerline{\psfig{figure=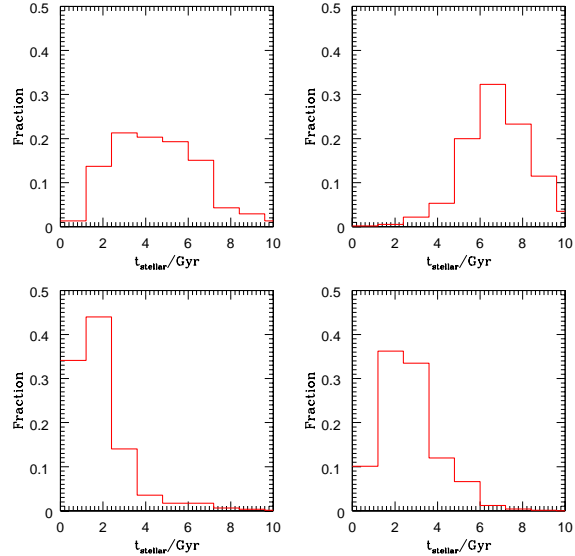,width=8cm,height=8cm}}
\caption{The stellar age, $t_{\rm stellar}$, for the different
spectral types: (bottom left) star-forming; (bottom right) composite;
(top left) AGN; (top right) unclassified. Typically $t_{\rm
stellar}\sim 1-4{\rm Gyr}$ for star-forming and composite galaxies,
whereas it is much larger ($t_{\rm stellar} > 5{\rm Gyr}$) for the
unclassified (passive elliptical) population. Interestingly the value
of $t_{\rm stellar}$ spans a wide range of values for the AGN
population.}
\label{fig:tstellar}
\end{figure}

\begin{figure}
\centerline{\psfig{figure=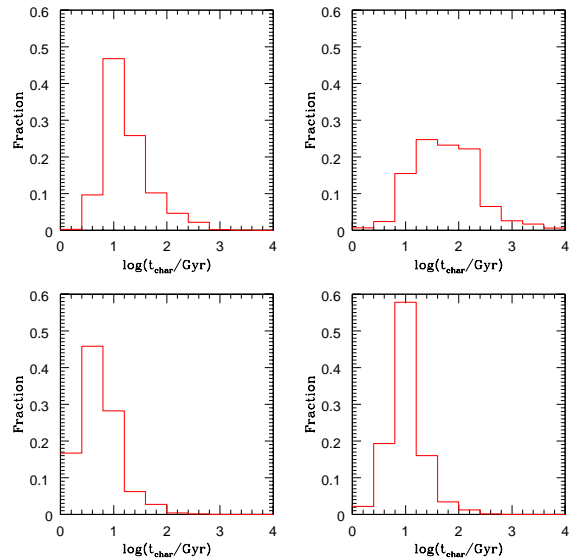,width=8cm,height=8cm}}
\caption{The characteristic age, $t_{\rm char}$, for the different
spectral types: (bottom left) star-forming; (bottom right) composite;
(top left) AGN; (top right) unclassified. Note that the scale is
logarithmic. The star-forming and composite populations have $t_{\rm
char}$ typically less than the age of the Universe, whereas for the
AGN and unclassified populations it is greater than the age of the
Universe.}
\label{fig:specific_spec}
\end{figure}

In Figs.~\ref{fig:tstellar} and \ref{fig:specific_spec} we have
plotted the fraction of the galaxies for which spectroscopy is
available as a function of $t_{\rm stellar}$ and $t_{\rm char}=m_{\rm
stellar}/{\rm SFR}$ for the different spectral types. The quantity
$t_{\rm char}$ is the time which would be required for the galaxy to
form all its stars at the rate they are presently being produced. Not
unexpectedly, we see that there is an increase in both quantities as
one goes through the spectral types from star-forming, through
composite and AGN, to the unclassified population. The star-forming
and composite galaxies have $t_{\rm stellar}\sim 1-4 {\rm Gyr}$ and
$t_{\rm char}$ typically less than the age of the Universe. The AGN
population have a very broad range of $t_{\rm stellar}$ and $t_{\rm
char}$ greater than the age of the Universe. The unclassified
population have $t_{\rm stellar}>5{\rm Gyr}$ and a broad range of
$t_{\rm char}$, typically $\sim 30-300{\rm Gyr}$. These properties are
compatible with our interpretation of this unclassified population, as
passive elliptical galaxies with a dominant red stellar population and
with the present day star-formation either being absent or involving a
very small fraction of the total stellar mass. We draw attention to
the fact that the AGN appear to be less prevalent amongst the galaxies
with the smallest stellar ages, possibly suggesting that radio-loud
activity tends to follow a burst of star-forming rather than being
coeval with it. A conclusion that activity in general (not just
radio-loud activity) follows a starburst was reached by \cite{Schawinski2007} based on their study of SDSS active galaxies. A detailed
discussion of the statistics of radio-jets and activity in nearby
galaxies is given in \cite{Kauffmann2008a}.

\begin{figure}
\centerline{\psfig{figure=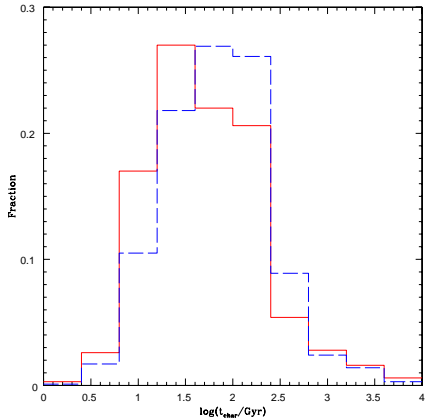,width=6cm,height=6cm}}
\caption{$t_{\rm char}$ for galaxies with $C_2>3.5$ and $b/a>0.6$, that is, those in the bottom two panels of Fig.~9. The solid line is the radio-quiet population with $t-r>-2.5$ and the dashed line is the radio-loud population with $t-r<-2.5$.}
\label{fig:specific}
\end{figure}

The quantities $t_{\rm stellar}$ and $t_{\rm char}$ can be used to the
help elucidate some of the properties of the interesting sub-samples we have
identified in the preceding sections. First let us consider the
sub-sample with $C_2>3.5$ and $b/a>0.6$ which was shown to have a
statistically significant correlation with $\Delta\alpha=90^{\circ}$
when the galaxy is relatively radio-quiet and has a uniform distribution of the
PA differences when the galaxy is radio-loud. For the most part the
distributions of photometrically measured and spectroscopically
inferred quantities are very similar for the radio-loud/quiet sub-samples. In
particular the distributions of $t_{\rm stellar}$ and $b/a$ 
are almost identical for the two
sub-samples. The most visibly different quantity is the distribution
of $t_{\rm char}$. Typically the distribution of $m_{\rm stellar}$ is
peaked at slightly larger values for the radio-loud sample, whereas
the SFR is peaked at a lower value. This combines to yield a
difference in the distribution of $t_{\rm char}$ between the two
samples which is presented in Fig.~\ref{fig:specific}. The median
value of $t_{\rm char}$ for the radio-quiet galaxies is around a
factor of 2 lower than that for the radio-loud galaxies.

\begin{figure}
\centerline{\psfig{figure=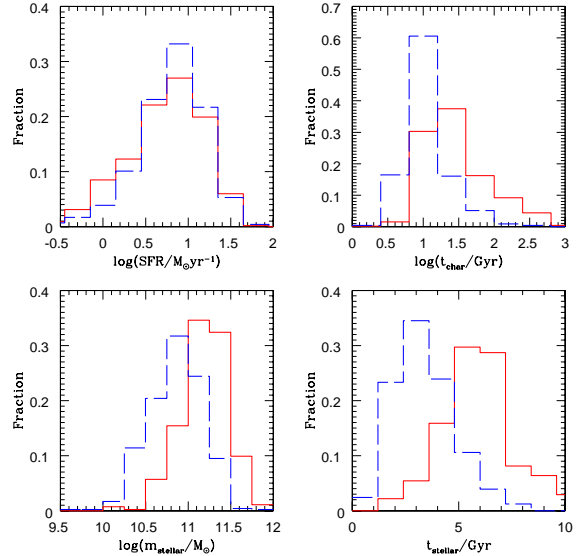,width=8cm,height=8cm}}
\caption{Properties of the AGN population split on the basis of $C_2$:
(bottom left) $M_{\rm stellar}$; (bottom right) $t_{\rm stellar}$; (top left) SFR in $M_{\odot}{\rm yr}^{-1}$; (top right) $t_{\rm
char}$. The solid lines are for $C_2>3.5$, whereas $C_2<3.5$ for the
dashed lines.}
\label{fig:per1agn}
\end{figure}

We have shown in Fig.~\ref{fig:histagn} that the AGN population can be
split using the parameter $C_2$ into a sub-sample which has a
significant correlation toward $\Delta\alpha=0^{\circ}$ ($C_2<3.5$)
and one with an almost uniform distribution of PA differences
($C_2>3.5$). The properties of the stellar populations of the AGN hosts when
split on the basis of $C_2$ are presented in Fig.~\ref{fig:per1agn}. We see
that the distribution of SFR is very similar for the two sub-samples,
but that objects with $C_2>3.5$ typically have a larger $m_{\rm stellar}$
implying that the $t_{\rm char}$ is typically larger. The sample with
$C_2<3.5$ has $t_{\rm char}$ sharply peaked around the age of the
Universe similar to that for the composite population, whereas the
distribution of $t_{\rm char}$ is much broader and peaked at around
$30~{\rm Gyr}$ for $C_2>3.5$ more in keeping with that of the
unclassified population. The distribution of $t_{\rm stellar}$ is also
interesting. For $C_2<3.5$ we find that $t_{\rm stellar}$ is typically
$<5{\rm Gyr}$ whereas for $C_2>3.5$ it is $>5{\rm Gyr}$.

\begin{figure}
\centerline{\psfig{figure=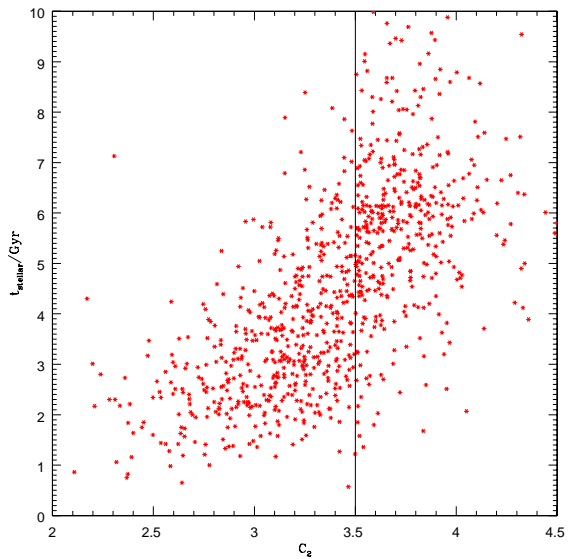,width=8cm,height=8cm}}
\caption{Scatter plot of $t_{\rm stellar}$ vs. $C_2$ for galaxies which were identified as being dominated by an AGN. Included also is the line $C_2=3.5$ which has been used to split these galaxies into a sub-sample biased toward $\Delta\alpha=0^{\circ}$ and one where the PA differences are uniformly distributed. There appears to be a strong correlation between the two quantities.}
\label{fig:corr1vage}
\end{figure}

The above results suggest that there could be a correlation between
$t_{\rm stellar}$ and the parameter $C_2$ for AGN.  In
Fig.~\ref{fig:corr1vage} we plot $t_{\rm stellar}$ against $C_2$.
There is a clear correlation; there is reasonably small scatter about
a line given by
\begin{equation}
{t_{\rm stellar}\over{\rm Gyr}}\approx 0.5(8 C_2-19).
\end{equation}
There is no hint of bimodality in this distribution. It is interesting
to note that star-forming and composite galaxies would typically
reside in the bottom left of this diagram ($t_{\rm stellar}<5{\rm
Gyr}$ and $C_{2}<3.5$) whereas the unclassified population would
reside in the top right ($t_{\rm stellar} > 5{\rm Gyr}$ and
$C_2>3.5$). The position angle differences for the AGN with young
stellar populations appear to share the same properties as the other
galaxies in that quadrant. On the other hand,  the older AGN appear not to
share the same properties as the other galaxies in their quadrant, though the numbers are small and the corresponding statistical uncertainties are large.

\subsection{Intrinsic alignment between the radio and optical structures}

An important question is the strength of the intrinsic alignment
between the radio and optical emission both for the star-forming
galaxies and the passive ellipticals. In both cases the distributions
of measured position angle differences are broadened by measurement
noise. The observed variance is due the intrinsic effect
$\langle(\Delta\alpha)^2\rangle_{\rm I}$ and measurement noise
$\langle(\Delta\alpha)^2\rangle_{\rm M}$, which we will assume are
independent and can, therefore, be added in quadrature
\begin{equation}
\langle(\Delta\alpha)^2\rangle=\langle(\Delta\alpha)^2\rangle_{\rm I}+\langle(\Delta\alpha)^2\rangle_{\rm M}\,.
\end{equation}
In what follows we will attempt estimate the value of
$\langle(\Delta\alpha)^2\rangle_{\rm I}^{1/2}$ based on a few simple
assumptions. Our intention is to give an idea of the level of
correlation rather than presenting a robust statistical analysis.

In addition to measurement noise, there can be projection
effects. These are likely to be negligible in the case of star-forming
galaxies which can be assumed to be disk-like with roughly co-spatial
radio and optical emission, but could be important in ellipticals
which can be triaxial \cite{Binney1978}. The possible effects of
triaxiality on the observed radio/optical position angle distribution
were simulated by \cite{Sansom1987}. Their simulations
show that, for oblate spheroidal galaxies, those with jets along the
minor axis should show the kind of alignments we observe, whereas
other configurations lead to much less well defined position angle
difference distributions; any triaxiality only serves to blur to the kind
of biases we observe. An important recent contribution has been
made by \cite{Padilla2008} who have analysed the shapes of
282203 elliptical galaxies from SDSS. They reach the conclusion that
the intrinsic shapes of ellipticals are consistent with them being
oblate spheroids (with the degree of flattening increasing with
decreasing luminosity). The implication of ellipticals being oblate is
that projection effects should not be significant and, therefore, we will
ignore them

The measurement error, quantified by
$\langle(\Delta\alpha)^2\rangle_{\rm M}$, can be reduced by focusing on
galaxies which are bright both in the radio and in the optical and, in
addition, have low values $b/a$. However, some caution is required
since it is possible that the correlation we are interested in is
intrinsically connected to the value of $b/a$ (or even the optical or 
radio flux); for example, galaxies
with low $b/a$ might have an intrinsically smaller dispersion in
position angle differences. We have found no evidence of this in the
case where the bias is toward $\Delta\alpha=0^{\circ}$ with all the
results for $C_2<3.5$ and star-forming galaxies being compatible with
measurement noise being dominant in the histograms when considered as
a function of $b/a$. The situation is more complicated in the case of
$C_2>3.5$ and passive ellipticals, where the strength of the
correlation appears to
improve as one increases the lower limit on $b/a$. This suggests that
the true correlation is stronger than we have observed. We will assume 
that we can measure $\langle(\Delta\alpha)^2\rangle_{\rm M}$ as a function
of $b/a$ for star-forming galaxies and that this holds
for elliptical galaxies.

\begin{figure}
\centerline{\psfig{figure=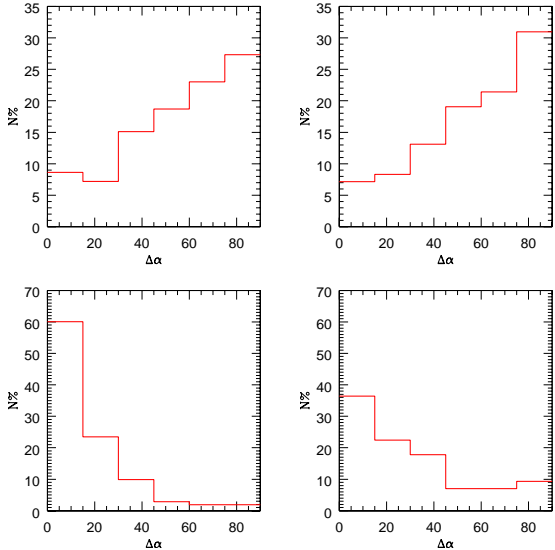,width=8cm,height=8cm}}
\caption{Histograms of the percentage of objects in each bin when the
sample is cut so as to maximize the number of objects with a
particular correlation. The bottom two plots are selected to have
$C_2<3.5$, $r<15$ and $S_{\rm int}>5{\rm mJy}$. In the plot on the
left, we have $b/a<0.5$, whereas on the right we have $b/a>0.6$. The
top two plots have $C_2>3.5$, $r<15$ and $S_{\rm int}>5{\rm
mJy}$. This time we plot those with $b/a>0.6$ on the left and $b/a>0.7$ on
the right.}
\label{fig:final}
\end{figure}

Again for objects with $C_2<3.5$, we have made cuts to the sample so as
to give the strongest possible results in terms of the percentage of
objects in each bin. We note that this is a somewhat different
philosophy to that used in the rest of the paper where we have
attempted to maximize the $\hat\chi^2$.  The results of this process
are presented in Fig.~\ref{fig:final}. We have imposed stringent cuts
of $r<15$, in order to improve the accuracy of determination of the
optical PA, and $S_{\rm int}>5{\rm mJy}$, to do the same for the radio
PA.  We will make the assumption that for $b/a<0.5$ and $C_2<3.5$ the
observed correlation is purely from the intrinsic scatter and that
measurement errors are negligible. By performing simulations with
different Gaussian dispersions, and ignoring the very few outliers, we
deduce that the dispersion seen in the case where $b/a<0.5$ is
compatible with a standard deviation of
$\langle(\Delta\alpha)^2\rangle^{1/2}_{\rm I}\approx 15^{\circ}$. This
is a conservative approach since by not explicitly taking into account
any radio position angles measurement errors, any intrinsic spread we
deduce will be an over estimate of the true value. Although the
existence of a correlation between the radio and optical PAs in these
systems may not be unexpected, the fact that it is so strong is
possibly a little more surprising.

Again for objects with $C_2<3.5$ we now consider the correlations when
$b/a>0.6$ (which is presented Fig.~\ref{fig:final}) and $b/a>0.7$. We
find that the histogram with $b/a>0.6$ is compatible with a total
dispersion of $\langle (\Delta\alpha)^2\rangle^{1/2}\approx
30^{\circ}$, and for $b/a>0.7$ it is compatible with $\approx
40^{\circ}$. Assuming that these galaxies are from the same parent
population as those with $b/a<0.5$ -- that is they have the same
intrinsic scatter -- and that the intrinsic effect is uncorrelated
with the effect of the axial ratio -- that is they can be added in
quadrature -- we can deduce $\langle(\Delta\alpha)^2\rangle_{\rm
M}^{1/2}$. We find that it to be $\approx 25^{\circ}$ for $b/a>0.6$
and $\approx 37^{\circ}$ for $b/a>0.7$.

For the $C_2>3.5$ population, essentially the passive ellipticals, we
have already shown that the correlation toward
$\Delta\alpha=90^{\circ}$ is strengthened as $b/a$ increases. The
histogram with $b/a>0.6$ is compatible with a total dispersion of
$\approx 45^{\circ}$. If we assume that the measurement error arising
from large axial ratios is the same as in the case of $C_2<3.5$, then
one can deduce that the combination of the intrinsic effect
($\langle(\Delta\alpha)^2\rangle_{\rm I}^{1/2}$), and any residual
effect due to projection, is $\approx 37^{\circ}$. For $b/a>0.7$, the
histogram is compatible with a total dispersion of $\approx
40^{\circ}$ and therefore one can deduce that the intrinsic plus
residual projection effects (assumed small if they are oblate
spheroids) give a dispersion of $\approx 15^{\circ}$. This is a
surprisingly small value, especially given that previous
investigations, albeit with much smaller numbers (for example, \cite{Sansom1987}), failed to produce a statistically significant
result. Physically the likely implication of such a small dispersion
is that we are dealing with well ordered systems in which the overall
shape of the host galaxy dictates the nuclear accretion axis and thus
the spin axis of the central black hole. The fact that the intrinsic
dispersion appears to be smaller in the more circular systems is again
unexpected.

\section{Discussion}

In the previous section we have identified and refined two
correlations between the optical and radio PAs; one is between the two
major axes ($\Delta\alpha=0^{\circ}$) and the other is between the
major and minor axes ($\Delta\alpha=90^{\circ}$). In the simplest
terms star-forming, predominantly spiral, galaxies have
$\Delta\alpha=0^{\circ}$ and passive ellipticals have a bias
towards $\Delta\alpha=90^{\circ}$.  We have tried to refine the
division between objects displaying the different types of behaviour
and in tables~\ref{tab:photo} and \ref{tab:spec} we present a summary
of the various cuts made on the data based on photometric and
spectroscopic properties respectively. We find that a new
composite parameter $C_2=0.262c+0.965(u-r)$ can be used to
discriminate better between different populations, with most objects
with $C_2<3.5$ having identifiable emission lines in their spectra,
and those with $C_2>3.5$ typically being passive ellipticals with no
identifiable lines. Use of the $C_2$ parameter improves the
statistical significance of the position angle difference
separation. The propensity to select passive ellipticals can be
increased by additionally requiring that $b/a>0.6$. Interestingly,
galaxies identified as being dominated by an AGN appear to straddle
the line $C_2=3.5$. In these, as in the galaxy population as a
whole, AGN with younger stellar populations and those with low axial 
ratios have $\Delta\alpha=0^{\circ}$, whereas those with older stars 
which are also more circular have a uniform distribution of PA differences.

\begin{table}
\begin{center}
\begin{tabular}{|c|cc|c|c}
\hline
Split & N & $\hat\chi^2$ & Bias & Fig\\
\hline\hline
All & 14302 & 117 & $0^{\circ}$ & \ref{fig:histall6}\\
\hline
$u-r<3$ & 8933 & 212 & $0^{\circ}$ & \ref{fig:histsplit1}\\
$u-r>3$ & 5369 & 3.6 & $90^{\circ}$ & \ref{fig:histsplit1}\\
$c<3$ & 8803 & 215 & $0^{\circ}$ & \ref{fig:histsplit1}\\
$c>3$ & 5499 & 2 & $90^{\circ}$ & \ref{fig:histsplit1}\\
\hline
$C_2<3.5$ & 7573 & 260 & $0^{\circ}$ & \ref{fig:histcorr}\\
$C_2>3.5$ & 6729 & 5 & $90^{\circ}$ & \ref{fig:histcorr}\\
$C_2>3.5$, $b/a<0.6$ & 1483 & 1.7 & $0^{\circ}$ & \ref{fig:histcorr}\\
$C_2>3.5$, $b/a>0.6$ & 5246 & 7 & $90^{\circ}$ & \ref{fig:histcorr}\\
\hline
$C_2>3.5$, $b/a>0.6$, $t-r>-2.5$ & 3441 & 11 & $90^{\circ}$ & \ref{fig:histsplit2}\\
$C_2>3.5$, $b/a>0.6$, $t-r<-2.5$ & 1805 & $<1$ & - & \ref{fig:histsplit2}\\
$C_2>3.5$, $b/a<0.6$, $t-r>-2.5$ & 1101 & 3.8 & $0^{\circ}$ & \ref{fig:histsplit2}\\
$C_2>3.5$, $b/a<0.6$, $t-r<-2.5$ & 382 & 1.6 & - & \ref{fig:histsplit2} \\
\hline\hline
\end{tabular}
\end{center}
\caption{Summary of the splits made on the data on the basis of the spectroscopic properties. $N$ is the number of objects satisfying the criteria and $\hat\chi^2$ is the $\chi^2$ per degree of freedom when compared to a uniform distribution.}
\label{tab:photo} 
\end{table}

\begin{table}
\begin{center}
\begin{tabular}{|c|cc|c}
\hline
Split & N & $\hat\chi^2$ & Bias\\
\hline\hline
All & 6053 & 56 & $0^{\circ}$ \\
Classified & 3752 & 114 & $0^{\circ}$ \\
Unclassified & 2301 & 3.8 & $90^{\circ}$ \\
Star-forming & 1593 & 97 & $0^{\circ}$ \\
Composite & 1158 & 39 & $0^{\circ}$ \\
AGN & 1001 & 3.1 & $0^{\circ}$ \\
\hline\hline
\end{tabular}
\end{center}
\caption{Summary of the splits of the data based on the classification using emission lines. Corresponding histograms are presented in Fig.~\ref{fig:histspec}. We used the same notation as in table~\ref{tab:photo}.}
\label{tab:spec} 
\end{table}

There is an obvious interpretation for the very strong bias toward
$\Delta\alpha=0^{\circ}$ in the galaxy population as a whole which is
that it arises from the subset of all galaxies in which there is
significant star-formation and where the radio emission comes, either
directly or indirectly, from supernova remnants associated with the
evolution of recently formed massive stars. There is little direct
evidence from the SDSS and FIRST data alone that the emission at the
two frequencies is accurately co-spatial but one should note that the
resolution of the radio images is a factor of $\sim$4 worse than that
of the optical images. Also single colour optical images are a
relatively inefficient way of distinguishing regions of active star
formation from the general stellar population. Hence there is
clearly scope for 1~arcsec resolution radio mapping and more detailed optical/infrared imaging of selected SDSS objects
to investigate in detail the correspondence between radio and optical
emission. \cite{Patel2009} are investigating in
detail the shapes of high redshift star-forming galaxies and their
associated emission using Hubble Deep Field data together with high
resolution VLA/MERLIN observations.

The bias towards $\Delta\alpha=90^{\circ}$ in the red, objects with
high concentrations is less strong but still highly statistically
significant. Since in these predominantly elliptical galaxies the
emission is almost certainly powered by mass accretion on to a central
disk/black hole system, the radio elongation we measure will be
related to the overall spin axis of that system. Our results confirm
those reported by \cite{Condon1991} who saw a strong trend for the
jet axes in UGC elliptical galaxies to be aligned with the optical
minor axes.  Simulations done by \cite{Sansom1987} for different
galaxy geometries show that one only gets the kind of strong minor
axis alignments that we observe for galaxies which are oblate
spheroids. We therefore interpret our results as implying that there
is a bias for the spin axis of the central engine to be aligned with
the minor axes of galaxies and also that these galaxies are
predominantly oblate spheroids. That elliptical galaxies are oblate
spheroids is consistent with the conclusions of \cite{Padilla2008} based on an analysis of galaxy shapes. The radio minor axis
alignment is something one might expect to see if the galaxies were
rotationally supported and the black hole accretion disk axes were
aligned with the overall galaxy rotation axes.
One should contrast what we see here in elliptical galaxies with what
seems to be a much more complex situation that appears to be common
in Seyfert galaxies where there is only a weak relationship between
jet axis and the axis of host galaxy disk (\cite{Kinney2000};
\cite{Gallimore2006}; \cite{Raban2009}).

The most remarkable thing that we find  about the
$\Delta\alpha=90^{\circ}$ bias in the elliptical galaxies is the fact
that strong alignment only seems to be present in the radio quieter
subset of these objects. We emphasise that we have not tried to
optimize our dividing line in radio loudness to maximize the
discrimination between the position angle properties. The division at
$t-r=-2.5$ separates the samples into sub-samples with a ratio in
numbers of around 2:1 and was not chosen {\it a priori} to have any
particular physical significance for the elliptical galaxy
population. However, our results suggest the division does have some
physical significance and is consistent with there being two types of
object within elliptical population exhibiting quite different
behaviour. We reiterate that the two radio populations we talk about here
have nothing to do with the traditional FR1/FR2 radio morphological
division between that occurs at around $t-r=-7$ not $t-r=-2.5$.

Based mainly on optical morphological and kinematic data, it has been
convincingly argued that there are indeed two sub-populations of
elliptical galaxies; there are those that are fairly round, are slowly
rotating and are generally massive, and there are those with more
elliptical isophotes, are fast rotating and are generally less massive
(see \cite{Emsellem2007}). These distinctions also appear to be
related to whether or not the galaxy has a core (see \cite{Faber1997};
\cite{Kormendy2009} and references therein for a comprehensive
discussion of elliptical galaxy properties). What we have found by
including radio position angle information suggests that such a split
may be evident down to the scale of the central black hole. In
radio-loud objects alignment of the central engine appears to be
independent of the overall shape of the host galaxy, while in the
radio-quiet population the engine knows about the galaxy shape. The
degree of radio-loudness is known to be a strong function of galaxy
mass (\cite{Best2005}) and the more massive galaxies tend to be of the
non-rotationally supported type. Thus it is very tempting to suggest a
simple picture in which the radio-quieter objects are predominantly
rotationally supported with matter on all scales sharing a common
rotation axis, while the louder ones are not rotationally supported
and have the complex kinematics often present in the rounder and most
massive galaxies (\cite{Emsellem2007}). However, the fact that amongst
the radio quieter population we see a stronger minor axis alignment in
the rounder galaxies does not have a natural explanation. Putting this
caveat aside, in the above picture having complex kinematics, perhaps
even a decoupled core, results in a more efficient radio-jet producing
engine, but one that can point in any direction with respect to
large-scale shape of the host galaxy.

To investigate the radio-quiet/radio-loud dichotomy further we have
looked for other differences in measured and inferred properties of
the two groups using the data from \cite{Kauffmann2003} and
\cite{Brinchmann2004}. The only significant differences we can find
are that median values of the stellar mass and $t_{\rm char}$ are a
factor of $\sim$2 lower in the radio-quiet population. These two
quantities are not independent since the star-formation rates in the
two types of galaxies are approximately the same and $t_{\rm char}$ is
the ratio of the stellar mass to $t_{\rm stellar}$. Since it is known
that the fraction of galaxies having radio luminosities above some
given limit increases rapidly with increasing host galaxy stellar mass
(\cite{Best2005}), the smaller average stellar mass for the
radio-quiet galaxies is only to be expected.  We have looked
explicitly to see if the degree of alignment we see depends on optical
luminosity (or stellar mass) and find that there is no obvious
dependence.  It is noteworthy that the lower characteristic age for
the stars amongst the radio-quiet population argues for a more
abundant supply of gas from stellar mass loss in these and this only
serves to emphasise the point that the efficiency with which the hot
gas available from the whole galaxy is converted into radio emission
must be significantly higher in the radio-loud objects. Of course
cold gas can be accreted too (see \cite{Kauffmann2008a}) but, being
passive elliptical galaxies, we expect objects in our sample to be in
the regime described by \cite{Kauffmann2008b} in which accretion is
limited by the supply of hot gas from evolved stars. Thus the physical
origin of apparent dichotomy remains a mystery but is probably related
in some way to the formation history of the galaxies since the effect
we observe depends on both the large-scale and small-scale
structure. In turn this could be related to the environment in which
the galaxy has evolved. For example, the radio-loud objects could live
in dense environments around groups and clusters, whereas the
radio-quiet objects could live in more quiescent environments. This
issue is presently under investigation (Battye et al., in
preparation).  An alternative hypothesis could be that the intrinsic
shape of the radio-loud galaxies is not oblate and hence projection
effects hide any intrinsic correlation which might exist.

The behaviour of the AGN population is also interesting. Galaxies
hosting active nuclei span the full range of galaxy types. Active
galaxies with young stellar populations have the radio and optical
axes aligned, albeit somewhat more weakly than in the case of
star-forming galaxies. This suggests that in these the radio emission
is dominated by that from supernova remnants which are correlated with
starlight.
The situation is less clear for galaxies with an older stellar
population hosting an AGN. These AGN appear to share nearly all of the
properties of the spectral unclassified population indicating that
they are in all other respects passive ellipticals. However, there is
no statistically significant trend for their radio emission to be
aligned with the host galaxy minor axis (Fig.~\ref{fig:histagn}), the
trend we see in the elliptical population as a whole. Perhaps this is
because of the relatively small numbers involved.

On the basis of the apparent correlation between $t_{\rm stellar}$ and
$C_2$ for AGN and the absence of any kind of bimodality in
Fig.~\ref{fig:corr1vage} it might be tempting to suggest that there
could be an evolutionary track from the the blue-extended region to
the red-compact region, as opposed to two separate populations, and
that AGN possibly represent disk galaxies in the process of turning
into passive ellipticals. Such a conclusion is supported by the work of \cite{Schawinski2007}. However, we think that this is probably an
over-simplification since the time-scale for the activity is likely to
be much shorter than that required to affect the transition from a
disk-dominated system into an elliptical.

There is one simple but quite significant consequence following from
the fact that we identify two different alignment behaviours in the
general elliptical galaxy population. It argues for the radio emission
that we observe being quiescent over long time-scales. It can be seen
from Fig.~\ref{fig:split} that a range of $t-r$ of 5 (a factor 100 in
flux density ratio) encompasses virtually all the objects in the
undivided population. Thus it would require variability of no more
than an order of magnitude to blur the distinction between the
alignment properties of the radio-quiet and radio-loud objects out of
existence. Plausible timescales on which aligned objects might be
transformed into a misaligned one are difficult to assess but probably
likely to be greater than the dynamical timescale, perhaps a few
$\times 10^8$~yr. Stability of the radio luminosity on this kind of
timescale would argue for a stable source of fuelling, most likely
mass loss from the stellar population consistent with the conclusion
recently presented by \cite{Kauffmann2008b}.

Finally, an original motivation for this study was to investigate what
might be found when very deep radio surveys are made with, for
example, the SKA in order to detect weak lensing.
Such studies are of interest because optical measurements may be
susceptible to a range of systematic effects. \cite{Chang2004} have
detected cosmic shear using the FIRST survey and claim that, in
principle, the radio should be less effected by systematics. More
conservatively, whatever they may be, the systematics are likely to be
different and hence uncorrelated. Therefore, correlating the signal
between the two wavebands will provide a cleaner result.
Our results suggest that the radio direction of elongation of
galaxies found in the very deep surveys required for weak lensing
studies will be the same as the optical with a dispersion of
$\approx 15^{\circ}$.

\section{Conclusions}

The results we have presented  show that consideration of the
relationship between the optical and radio position angles can lead to
a number of interesting inferences. The power of these results has
been significantly enhanced by the rich and varied astronomical
information available from the SDSS. We believe that our analysis
introduces a new slant on the very significant knowledge of galaxy
evolution which has already come from the SDSS. We have shown that
galaxies with young stellar populations which are forming stars at a
significant rate predictably have aligned optical and radio axes with
a dispersion of around $15^{\circ}$. Much fainter versions of such
star-forming galaxies will be useful for future weak lensing studies
and should complement optical studies of the same fields.

Perhaps more surprisingly we have found a correlation between the
radio major axis and the optical minor axis in radio-quiet passive
ellipticals which, once we take into account the range of noise and
projection effects related to the orientation, have a similarly low
dispersion to that seen in the star-forming galaxies. Such a
correlation has been long sought (for example, \cite{Mackay1971}) because
of its potential to tell us about the intrinsic shapes of the galaxies
and the relationship between distribution of stars and the engine that
generates the radio jets. Our results argue strongly in favour of
there being a population of oblate spheroidal galaxies with radio jets
aligned with the stellar minor axes. We find, in fact, evidence for
two populations of passive ellipticals. Those which are relatively
radio-quiet show strong radio alignment with the optical minor axis
while the radio louder population has no apparent correlation between
the optical and radio emission. We tentatively identify our two
populations with those independently found from optical photometric
and kinematic studies of ellipticals, the rounder non-rotating and
massive galaxies, and the more disky, rotating and less massive
ones. The rotationally supported disky galaxies will naturally be
oblate and have a single well-defined axis. On the other hand, the
slowly rotating galaxies may have complex kinematics with the jet axes
misaligned with the main stellar distribution. Jets may be bent as
they propagate and triaxiality will give rise to projection
effects. All this can go some way towards explaining the lack of
radio/optical alignments in the radio louder population. There is
clearly great scope for further detailed optical and radio
observations of these relatively radio-weak elliptical galaxies.

\section*{Acknowledgements}

\noindent We have made extensive use of the SDSS and the galaxy
properties derived from SDSS by the MPA-Garching group
(http://www.mpa-garching.mpg.de/SDSS/DR4/Data/sfr
catalogue.html). Funding for the SDSS and SDSS-II has been provided by
the Alfred P. Sloan Foundation, the Participating Institutions, the
National Science Foundation, the U.S. Department of Energy, the
National Aeronautics and Space Administration, the Japanese
Monbukagakusho, the Max Planck Society, and the Higher Education
Funding Council for England. The SDSS Web Site is
http://www.sdss.org/. We have also used the FIRST radio data obtained
with the NRAO VLA. The National Radio Astronomy Observatory is a
facility of the National Science Foundation operated under cooperative
agreement by Associated Universities, Inc.  This research has made use
of the NASA/IPAC Extragalactic Database (NED) which is operated by the
Jet Propulsion Laboratory, California Institute of Technology, under
contract with the National Aeronautics and Space Administration. We
thank Rob Beswick for useful discussions about the properties of
star-forming galaxies and Shude Mao for constructive suggestions and
also Mark Birkinshaw, Joan Wrobel and Anne Sansom for valuable
comments on an earlier draft of the paper.

{}
\end{document}